# Discounting with Imperfect Collateral

Lou Wujiang[1]

Initial draft Jan 18, 2017; Updated 6/11/2017.

**Abstract**

Cash collateral is perfect in that it provides simultaneous counterparty credit risk protection and derivatives funding. Securities are imperfect collateral, because of collateral segregation or differences in CSA haircuts and repo haircuts. Moreover, the collateral rate term structure is not observable in the repo market, for derivatives netting sets are perpetual while repo tenors are typically in months. This article synthesizes these effects into a derivative financing rate that replaces the risk-free discount rate. A break-even repo formulae is employed to supply non-observable collateral rates, enabling collateral liquidity value adjustment (LVA) to be computed. A linear programming problem of maximizing LVA under liquidity coverage ratio (LCR) constraint is formulated as a core algorithm of collateral optimization. Numerical examples show that LVA could be sizable for long average duration, deep in or out of the money swap portfolios.

## 1. Introduction

When we speak of securities financing, stocks and bonds come to mind, but never options, futures, and other derivatives. Repurchasing (repo) an option? To this date, there is no derivatives financing market. On the other hand, private financing of derivatives has become an increasingly dominant market trend. By private financing, we mean derivatives collateralization by cash or cashable securities that are exchanged for cash in the repo market. Credit support annex (CSA) and central counterparty (CCP) margining agreements effectively create a private derivatives financing market, where securities collateralized derivative assets can hurl in cash just as Treasury bonds do.

---

[1] The views and opinions expressed herein are the views and opinions of the author, and do not reflect those of his employer and any of its affiliates. An earlier version is titled as "Pricing non-cash collateralized derivatives and collateral optimization with liquidity value adjustment"

Such a private market has been on a rapid rise following the post-crisis redesign of the over-the-counter (OTC) derivatives market, as derivatives collateralization and margining become the main counterparty credit risk mitigation tool. It can be either bilateral between two counterparties or unimodal when parties clear through a CCP. On the CCP clearing front, firms post initial margins to CCP in addition to bi-way variation margins. On the bilateral front, new margin rules issued by the Basel Committee on Banking Supervision (BCBS) and the International Organization of Securities Commissions (IOSCO) for non-centrally cleared OTC derivatives and certain security financing transactions have phased in since September 2016. For the remaining out of the scope trade population, i.e., uncollateralized derivatives trades with non-covered institutions under BCBS-IOSCO, dealer's hedge with other dealer banks necessitates collateral flow.

As is a private market, its financing rate is not observable. In fact, the derivative financing rate we speak of is an artifact for derivative fair value, better known as the discount rate. In general, it is expected to deviate away from the risk free discount rate. Johannes and Sundaresan (2007) equate a swap under cash collateralization to a portfolio of futures contracts, and find empirical evidence that links cost of collateral or *costly collateral*[2] to swap rates. The daily MTM results in a stochastic collateral income or dividend, which is built as a convenience yield into the discount rate, superseding the usual Libor rate, the pre-crisis *de facto* standard for the risk free discount rate. The surge in the Libor OIS spread during the credit crunch of 2007-2008 quickly forged a consensus that fully collateralized swaps need to be discounted instead at the overnight indexed swap (OIS) curve.

Johannes and Sundaresan (2007)'s analysis is typical of the early risk neutral, reduced form approach that discounts swap cash flow at the risk-free rate conditional on no-default. Piterbarg (2010) takes a different approach in building the cost of collateral into the Black-Scholes-Merton pricing framework. The cash collateralized portion of the option MTM could earn a CSA contractual collateral rate, and the remaining uncollateralized portion earns the bank's own unsecured rate. The derivative's financing cost is a weighted sum of the cash collateral rate and the unsecured rate. Piterbarg (2012) shows that when the real world measure is changed to a new, equivalent risk neutral measure, (fully) collateralized assets return and thus should discount at their collateral rates rather than the usual risk-free rate.

---

[2] Costly collateral refers to the difference of the interest rate rebated for posted cash collateral from its acquisition cost. If the secured party reinvests the cash in Treasury bills and passes back earned interest, while the pledging party gets cash from the Fed funds window, there is a net cost as bills have a known liquidity premium, therefore having an interest rate lower than the Fed funds rate or GC repo rates.

For non-cash collateralized trades, Hull and White (2014) perceives that the collateral rate for securities collateral should be the risk neutral expected return, irrespective of haircuts. Others, e.g., Brigo, Liu, Pallavicini, and Sloth (2014) and Burgard and Kjaer (2013), tend to agree that the repo rate for the securities should be used. The excess return of the collateral rate over the risk free rate results in collateral valuation adjustment (ColVA). The exact nature and characteristics of the collateral rates are not provided, as they are assumed to be exogenously given, presumably by a repo financing desk that knows how to determine collateral rates of different collateral asset types for the entire duration of the derivatives netting set.

Note that non-segregated cash collateral is perfect collateral in a sense that it provides the same amounts of counterparty credit risk protection and derivatives funding simultaneously. Securities are imperfect collateral[3] in that the differences in CSA haircuts and repo haircuts create mismatched protection and funding amounts. CSA haircuts, e.g., 5% for 10 year US Treasury notes, are pre-agreed and static. Upon receiving securities, the secured party has to turn to the repo market to access cash. Repo haircuts, however, are known to be dynamic and procyclic, lower in an expansion cycle and higher in a contracting cycle.

A related issue with imperfect collateral is tenor mismatch. A netting set is essentially perpetual, while the repo market, however, is known for its short tenors of typically 3 months, rarely one year or beyond. Adding to the complexity is that repo rates are intimately related to haircuts and counterparty credit.

---

[3] Fujii and Takahashi (2013) study asymmetric and imperfect collateralization due to CSA's intricacies, unrelated to CSA and repo haircut differences.

This article's main contribution is to take collateral imperfection into account in derivatives pricing. The protection and funding mismatches due to CSA and market haircut differences are captured in a single derivatives financing or discount rate. This results in a neat formulation of the total valuation adjustment, which can be decomposed into credit, funding, and collateral adjustments, and a refined liquidity value adjustment (LVA). Collateral rates consistent with a newly developed repo pricing methodology are supplied, allowing realistic collateral financing costs to be computed via LVA. We also explore using LVA as a quantitative measure to aid collateral optimization, a critical business process that major financial institutions perform routinely and largely on a qualitatively basis.

**2. Imperfect Cash Collateral**

Cash collateral is only imperfect when segregated, as credit protection is provided at the cash amount, but no part of it is used to fund the derivatives. This becomes an intermediate case between reused cash collateral that provides both protection and funding, and uncollateralized case that offers neither protection nor funding. Unlike reused cash collateralized trades, pricing of uncollateralized trades has seen different methodologies and debates. At the center is funding valuation adjustment (FVA): the dealer bank's unsecured funding cost or benefit on an uncollateralized (thus unfunded) derivative trade. Banks' FVA models (e.g. Burgard and Kjaer, 2011) could result in fair values which depend on the pricing agent's own funding curve and violate the law of one price.

Hull and White (2014) show that a dealer bank's funding of an uncollateralized asset would incur a debt valuation adjustment (DVA) that offsets its FVA cost to the extent that only the dealer's (liquidity) basis contribution is left. They further propose to replace it with an average of dealer banks' basis (e.g. bond CDS basis) such that the law of one price remains intact. Anderson, Duffie and Song (2016) prove that FVA should land in the firm's equity rather than its balance sheet. The fair value is unaffected, although traders may incorporate it into bid/ask to recoup value for equity holders. Lou (2016a) examines the industry standard FVA setup where an uncollateralized customer trade is hedged back-to-back with another dealer under full bilateral CSA, and points out that interest rate risk, e.g., IR01, is not hedged. Correction of the IR01 leakage by means of dynamic swap notional leads to the same liability-side pricing formulae of Lou (2015).

Specifically, the following partial differential equation (PDE) is obtained for $V$ -- the fair value of an uncollateralized stock option between parties B and C,

$$\frac{\partial V}{\partial t} + (r_s - q)S\frac{\partial V}{\partial S} + \tfrac{1}{2}\sigma^2 S^2 \frac{\partial^2 V}{\partial S^2} + r_b V^- - r_c V^+ = 0 \tag{1}$$

where $S$ is the stock price, $q$ stock dividend yield, $r_b(t)$ and $r_c(t)$ B's and C's bond interest rates respectively, $r_s(t)$ stock's financing rate, $\sigma$ volatility.

The risk neutral pricing formulae under $Q$-measure for an option with a terminal payoff function $H(T)$ is extended via a switching discount rate,

$$\begin{aligned} V(t) &= E_t^Q [e^{-\int_t^T r_e du} H(T)], \\ r_e(t) &= r_b I(V(t) \leq 0) + r_c I(V(t) > 0). \end{aligned} \tag{2}$$

The derivatives financing rate $r_e$ is the liability-side's unsecured rates. The (total) counterparty risk adjustment (CRA), denoted by $U$ as the difference between the risk-free derivative price $V^*$ and $V$, is precisely given by,

$$U = E_t[\int_t^T (r_e - r)V^*(s)e^{-\int_t^s r_e du} ds] \tag{3}$$

This formulae is interesting for it shows $r_e$ as both the discount rate -- appearing in the exponent, and the financing rate -- appearing in the funding spread applied to $V^*$. A decomposition of the spread $r_e - r$ into a default risk premium and the funding basis leads to a coherent form of CVA and FVA, which avoids overlapping DVA and FVA, is balance sheet neutral, conforms to the law of one price, and is thus better suited for fair value purposes.

For a partial cash collateral $L_t$ earning interest rate $r_L$, $V-L$ is the uncollateralized amount, and the LSP PDE is shown to be

$$\begin{aligned} &\frac{\partial V}{\partial t} + (r_s - q)S\frac{\partial V}{\partial S} + \tfrac{1}{2}\sigma^2 S^2 \frac{\partial^2 V}{\partial S^2} - rV \\ &+ (r_b - r)(V - L)^- - (r_c - r)(V - L)^+ - L(r_L - r) = 0, \end{aligned} \tag{4}$$

With full collateralization, the derivative funding cost in the PDE returns to the cash collateral funding rate $r_L$, same as in Piterbarg (2010).

It is worth noting that in the derivation leading to PDE (1) and (4), the cash collateral is allowed to comingle with the economy, a typical setup for variation margin. Smaller end users are allowed to elect to have their collateral segregated, even for variation margin. Comingled cash is usable for the economy's general purposes, and economically serves both purposes of credit mitigation and derivatives funding. As a special case of imperfect collateral, segregated cash collateral only provides credit mitigation but not funding. In the section below, we follow Lou (2015) to extend equation (1) for segregated cash collateral.

**2.1 Segregated cash collateral PDE**

Consider party B (a hypothetical bank) and party C (a customer) enter into an option trade with the bank dynamically hedging the option. Both parties have access to a liquid corporate bond market, primary or secondary, exogenous to the simple option economy and stock $S$ is financed in the repo or security lending market. When party B has a positive exposure to party C (the derivative is a receivable to B), C posts $L_s$ amount of cash collateral segregated in a separate account, $L_s \leq V$. Income earned on collateral is directly returned to C which B has neither control nor security interest. At the time of C's default, cash will be released from the segregated account to B to cover any close-out payment.

Prior to a default termination, B has to find a way to finance its derivatives. Following Lou (2015), the unsecured amount $W_t$, $W=V- L_s \geq 0$, is financed by the liability-side. $W_t = W_t^+ - W_t^-$, $W_t^+$ is the cash amount deposited or posted by party C to B that pays C's cash rate $r_c(t)$, and $W_t^-$ is by B to C earning B's cash rate $r_b(t)$.

The wealth equation of a long option economy from party B's perspective is
$$\pi_t = M_t + (1-\Gamma_t)(V_t - W_t - N_t - \Delta_t S_t + L^s), \qquad (5)$$
where 1-$\Gamma$ is the joint survival indicator, $L^s$ a stock lending account with zero haircut on $\Delta$ shares of stock, $L^s=\Delta S$. The stock short sale proceed is deposited with the stock lender who pays rebate interest at the rate of $r_s$. $M_t$ is the bank account that earns the risk free deposit rate $r$. $N_t$ is B's debt account that issues short term rolled debt at par rate $r_N(t)$, $r_N(t) \geq r(t)$. The account could be secured by the remaining asset of the economy and could have recourse to the bank.

Cash flow on the segregated cash account does not enter the economy's financing equation[4]. Pre-default self-financing equation then includes dynamic hedging, stock financing, and collateral rebalancing cashflow detailed as follows,

$$dM_t = rM_t dt + (1-\Gamma_t)[d\Delta_t(S_t + dS_t) - \Delta_t Sqdt + dN_t - r_N N_t dt \\ + dW - r_c W^+ dt + r_b W^- dt - dL^s + r_s L^s dt] \quad (6)$$

At $t=0$, the wealth reduces to $\pi_0 = M_0 + V_0 - W_0 - N_0$. Since $V-W\geq 0$, we set $M_0=\pi_0=0$ and $N_0=L_0\geq 0$. This says that the economy borrows a start cash of $L_0$, exactly the segregated amount.

For $t>0$, keeping the pre-default wealth to 0 leads to $N_t = L_s$ and $M_t=0$. Now suppose that party C defaults at time $\tau$, unwinding the stock financing gets back cash amount of $\Delta S$, buying back stock hedges pays cash amount of $\Delta S$, and the segregated cash collateral account pays $L_s$. The unsecured part of $V$ is set-off by the liability-side deposit $W$ (Lou 2015). The net inflow is $L_s$, exactly same as $N$, so the debt account can be cleared without a loss. There is no jump in cash flow at default and equation (6) is indeed the financing equation covering both pre-default and post-default. This is also true when party B defaults first.

Differentiating equation (5) and plugging in equation (6), we obtain,

$$d\pi_t - r\pi_t dt = (1-\Gamma_t)[dV_t - rV_t dt - (r_c - r)W^+ dt + (r_b - r)W^- dt \\ - \Delta_t(dS_t - (r_s - q)S_t dt) - (r_N - r)N_t dt] \quad (7)$$

Without loss of generality, write $r_N(u) = \mu_b I(V(u) \leq 0) + \mu_c I(V(u) > 0)$. Plugging into equation (7) and setting $\pi_t = 0$ result in,

$$dV_t - r_N V_t dt - (r_c - \mu_c)W^+ dt + (r_b - \mu_b)W^- dt - \Delta_t(dS_t - (r_s - q)S_t dt) = 0 \quad (8)$$

Now apply Ito's lemma, assume delta hedge under the usual geometric Brownian motion stock price ($dS=\mu Sdt+\sigma SdW$), and set $dt$ term to zero, we obtain the following,

$$\frac{\partial V}{\partial t} + (r_s - q)S\frac{\partial V}{\partial S} + \tfrac{1}{2}\sigma^2 S^2 \frac{\partial^2 V}{\partial S^2} - r_c V^+ + r_b V^- + (r_c - \mu_c)L_s^+ - (r_b - \mu_b)L_s^- = 0 \quad (9)$$

---

[4] This is obvious when a third party custodian is hired to manage the segregation account.

The first five terms are same as the LSP PDE (1). The segregated cash collateral enters into the last two terms. The meaning of rates $\mu_b$ and $\mu_c$ becomes clear when we consider full cash segregation, $L_s=V$. Equation (9) becomes

$$\frac{\partial V}{\partial t} + (r_s - q)S\frac{\partial V}{\partial S} + \tfrac{1}{2}\sigma^2 S^2 \frac{\partial^2 V}{\partial S^2} - \mu_c V^+ + \mu_b V^- = 0 \qquad (10)$$

Since the debt account $N_t$ amount is fully repaid upon default termination, charge rates $\mu_b$ and $\mu_c$ should not contain default risk premium. They reflect bond market liquidity and other non-credit, market structural factors (Hull and White 2014), and can be referred to as the issuer liquidity rates.

A firm's liquidity rate $\mu$ might not be directly market observable but can be implied from the cash bond market and CDS market via a basis trade. Assuming that C's CDS is cleared through a CCP at a short term par CDS premium $x$, $x$ is free of counterparty risk. Without considering gap risk, the basis trade is not subject to C's default risk, so its no-arbitrage pricing rate has to be $r_c - x$, i.e., $\mu_c = r_c - x$. $r_c - x - r$ is then the bond CDS basis (or funding basis). Under zero-recovery assumption, $\mu_c = r_c - \lambda_c$, where $\lambda_c$ is C's default intensity.

$r_b - \mu_b$ and $r_c - \mu_c$ stand for B's and C's default risk premiums respectively. The sixth term in equation (9) applies when C has to post. Since $r_c - \mu_c > 0$, it is a protection benefit, or gain due to protection afforded under the segregated collateral account. The last term with $r_b - \mu_b$ applies when B is posting, reflecting a loss of value due to providing protection to counterparty, as compared to nothing provided when uncollateralized. PDE (9) satisfies the law of one price[5].

**2.2 Discount rate representation**

---

[5]The liquidity rate's fit for fair value purposes can also be illustrated by a market equilibrium. Suppose that the option is an asset to B. B issues a credit linked note (CLN) referencing the option and its segregated cash collateral, linked to C's credit, in the amount of $L_s$ with $r_d$ being its price. Because $r_d$ directly impact derivatives pricing, party C has vested economic interest in $r_d$. If $r_d$ is too high, for example, and C finds his own funding cost is lower, C would step in to buy the CLN, willing to pay at his own funding rate, thus lowering $r_d$. If $r_d$ is already low in the market place, C benefits, but B would want to have a claim of that benefit, essentially driving up $r_d$ back to C's liquidity rate in the equilibrium.

We already know that with comingled full cash collateral providing both credit protection and funding, the correct discount rate to use is the risk-free rate. For fully uncollateralized trades with neither protection nor funding, the proper discount rate, according to Lou (2015), is the liability-side's senior unsecured rate. Segregated cash collateral provides credit protection but not funding, and the appropriate discount rate shown in equation (10) is the liability-side's liquidity rate, i.e., the risk-free rate plus the liquidity basis.

Write the segregated amount in a proportion form for $V \neq 0$,

$$\eta_c = \frac{L_s^+}{V^+}, \eta_b = \frac{L_s^-}{V^-}$$

Then PDE (9) becomes

$$\frac{\partial V}{\partial t} + (r_s - q)S\frac{\partial V}{\partial S} + \tfrac{1}{2}\sigma^2 S^2 \frac{\partial^2 V}{\partial S^2} - r_e V = 0, \qquad (11)$$
$$r_e = (r_c(1-\eta_c) + \eta_c \mu_c)I(V>0) + (r_b(1-\eta_b) + \eta_b \mu_b)I(V \leq 0)$$

The Feynman-Kac formulae would lead to the same expectation formulae as in equation (2) with the discount rate $r_e$ replaced by (11).

We could introduce a flag $\chi$, 0 meaning segregated, 1 comingled. Since non-segregated cash returns at rate $r_L$, the effective discount rate for the PDE can be rewritten to accommodate both cases.

$$r_e = r_{ec}I(V>0) + r_{eb}I(V \leq 0), $$
$$r_{eb} = r_b(1-\eta_b) + \eta_b((1-\chi_b)\mu_b + \chi_b r_L)), \qquad (12)$$
$$r_{ec} = r_c(1-\eta_c) + \eta_c((1-\chi_c)\mu_c + \chi_c r_L))$$

For party C, its effective rate $r_{ec}$ is a linear combination of the unsecured rate $r_c$, the liquidity rate $\mu_c$, and the cash rate $r_L$, each applied to the unsecured (and unfunded) portion $1-\eta_c$, secured but unfunded portion $(1-\chi_c)\eta_c$, and $\chi_c\eta_c$ the remaining funded portion of the fair value $V^+$. $r_{eb}$ can be understood similarly.

In our construct, the only nonlinearity appears in the effective discount rate, unlike Brigo et al (2014) where a separate non-linear valuation adjustment is introduced in their backward SDE to correct an overlap in value adjustments. Synthesizing into one derivative financing rate has its advantages. For example, it is now obvious that party C's zero coupon bond will be priced at C's senior unsecured rate $r_c$, a model consistency test not satisfied by all. Also as demonstrated in Lou (2016a), a Monte Carlo simulation with regression procedure can be developed focusing on the convergence of the finance rate. Valuation adjustments correspond to its decomposition and can be computed seamlessly thereafter.

This concept of an effective derivative financing rate in the presence of counterparty credit risk and collateral is also evident in Hull and White (2014), where $r_d$ is shown for various scenarios of collateralization and value adjustments. It is not seen in Burgard and Kjaer (2011), although their main result 1 (when the close-out uses risky market value) can be easily converted to yield $r_{ec} = (1-R_c)(r_c - r) + r_F$ and $r_{eb} = (1-R_b)(r_b - r) + r$ where $R$ is the recovery rate, $r_F$ equals $r$ if the derivative can be repo-ed and $r_{eb}$ otherwise. Brigo et al (2014) and Anderson, Duffie, and Song (2016) do not present such a form. Duffie and Huang (1996) has a similar switching discount rate in the swap fair value that equals to the sum of the risk free rate and the CDS spread. Lou (2016a) shows that adding a carry cost term can result in the same formulae as eqt. (2).

## 3. Non-cash Collateral

For non-cash collateral, the CSA normally stipulates that all cash income generated from the posted securities will be returned to the posting party. Unlike cash collateral which does not incur cost on the secured party, converting securities to cash by means of repo involves cost, as the repo rate is higher than the cash rate. Theoretically, such a cost could be deducted from the income earned before it gets passed back to the pledger. This is however not what the CSA provides for. Collateral repo cost has to be incorporated in derivatives pricing.

### 3.1 Non-cash collateral rehypothecation

Upon receiving reusable securities from a pledger, the secured party can pledge or sell them in a repo transaction to raise cash or rehypothecate in a separate transaction with a different derivatives counterparty. The latter is economically the same as if the secured party transforms the securities to cash in a repo transaction and subsequently pledges the cash to the other party, so we can simply assume that securities are always repo-ed out for cash.

Suppose party C posts to party B with securities traded at a haircut $h_p$ and a repo rate $r_p$. At the same time, a haircut $h_c$, not necessarily same as $h_p$, is applied per CSA. Let $n$ denote the number of units of securities posted, and $B_t$ the price of posted securities. The market value of the securities is $nB_t$. Credit protection under the CSA is then $L=\min(nB_t(1-h_c),V^+)$, while the maximum cash equivalency or funded amount $L_c = nB_t(1-h_p)=L(1-h_p)/(1-h_c)$.

Let's begin with a special case when $h_p=h_c$, then $L=L_c$. As securities are comingled, $\chi=1$. The repo financing rate $r_p$ now replaces the collateral rate $r_L$ in equation (12) so that the effective discount rate is modified as follows

$$\eta_c = \min(1, \frac{nB_t(1-h_c)}{V^+}), \eta_b = \min(1, \frac{nB_t(1-h_c)}{V^-}) \tag{13}$$
$$r_e = (r_c(1-\eta_c)+\eta_c r_p)I(V>0)+(r_b(1-\eta_b)+\eta_b r_p)I(V \leq 0)$$

For full collateralization, we have $\eta_b=\eta_c=1$, then $r_e = r_p$.[6]

When $h_p \neq h_c$, the credit protection amount $L$ and the cash equivalency amount $L_c$ do not equal. Consider a full collateralization case $L=V$: when $h_p>h_c$, $L_c$ is less than $V$, so B has deficient fund $V-L_c$; when $h_p<h_c$, $L_c>V$, party B then has excess fund $L_c-V$.

The deficient fund resembles a segregated cash amount as it is protected and bears no credit risk, so C's liquidity rate applies. With an excess fund amount of $L_c-V$, it is debatable whether to include its share of the repo cost. The bank may choose not to rehypothecate the portion of collateral that corresponds to the excess fund, thus avoiding the cost. Or the bank reuses the excess cash for other purposes, but then it should pay for it. It is thus commercially unreasonable to pass through the cost due to the excess funding, so the cost basis to be charged through the valuation PDE should be capped at $V^+$, i.e., $L_c = \min(V^+, nB_t(1-h_p))$.

With partial collateralization, we adopt the same pricing scheme: the deficiency fund created by $h_p>h_c$ can be charged at C's liquidity rate, and the excess fund created by $h_p<h_c$ is not utilized and not charged of any cost or passed of any benefit. Incorporating these considerations, the effective discount rates for party B and C are revised to

$$\chi = 1 - (\frac{h_p - h_c}{1 - h_c})^+,$$
$$r_{eb} = r_b(1 - \eta_b) + \eta_b((1 - \chi_b)\mu_b + \chi_b r_p)), \tag{14}$$
$$r_{ec} = r_c(1 - \eta_c) + \eta_c((1 - \chi_c)\mu_c + \chi_c r_p)),$$

---

[6] Our repo rate notation does not distinguish party B from C as repo borrowers. We could write $r_p = r_{pC}I(V>0)+r_{pB}I(V\leq 0)$, where $r_{pC}$ and $r_{pB}$ denote the repo rates with recourse to C and B respectively.

In the above, we assume that the pricing of the collateral securities and the derivatives can be independently conducted, i.e., there is no wrong way risk. When securities other than Treasuries are eligible, CSAs commonly set out eligibility criteria to restrict low quality and tangled securities. Under robust front office due diligence, risk management monitoring, and control processes implemented, the securities posted do not incur specific wrong way risk with regards to the counterparty and the derivatives products. The existence of reasonably sized haircuts further weakens general wrong way risk (Lou 2016c) to such an extent that it can be safely ignored to the first order. As such, there is no need to enlist $B_t$'s dynamics.

### 3.2 Initial margin collateral

The discount rates derived above are for variation margins. Initial margin (IM) is different because collateral segregation is mandated in CCPs and under BCBS-IOSCO rules. Also non-cash collateral dominates IM posting.

Trivially, segregated securities do not provide any funding benefit, so $L_c=0$. CSA haircut still applies to the protection amount, i.e., $\eta_c = \min(1, \frac{nB_t(1-h_c)}{V^+})$. The discount rates in equation (14) remain valid by simply letting $\chi=0$.

There could be a separate issue in that the existing CSA haircut is no longer sufficient, comparing to a would-be new CSA haircut. This could happen for instance when the eligible securities classes have experienced a worse stress after the CSA was signed. The difference can be easily treated as unsecured exposure or the bank can apply the new haircut for pricing purposes. (Of course, legally, party C still posts under the existing CSA haircut.) For now, we assume that the existing CSA haircuts are sufficient to mitigate the credit risk.

Parties can't draw any funding benefit from each other's posted IM collateral. It is also impractical to charge one's IM funding cost in the form of margin value adjustment (MVA) over to the other party, unless one is in market making capacity. It is possible however to charge MVA in a non-covered customer trade. Since the dealer has the option of posting cash and non-cash, he could price the MVA on the prudent side by assuming cash collateral. If such is the case, IM haircuts don't need to enter fair value calculation.

**4. Liquidity Value Adjustment**

As the fair value $V_t$ solved from PDE (11) fully incorporates counterparty credit risk and derivatives funding cost, it does not need to be adjusted. With regard to the (counterparty) risk-free value $V_t^*$ which satisfies PDE (11) with $r_e = r$, the total valuation adjustment (XVA) is trivially the difference of $V_t$ and $V_t^*$. Let $U = XVA = V^* - V$, then $U$ is governed by

$$\frac{\partial U}{\partial t} + (r_s - q)S \frac{\partial U}{\partial S} + \tfrac{1}{2}\sigma^2 S^2 \frac{\partial^2 U}{\partial S^2} - r_e U + (r_e - r)V^* = 0 \tag{15}$$

Noting $U_T = 0$, application of Feynman-Kac theorem immediately leads to the same XVA formula as in equation (3). This is of course mostly a representation rather than an actual solution as the switching rate is coupled with $V$. Nonetheless, a breakdown of XVA can be obtained, for example, $U = CVA - DVA + CFA - DFA + LVA$,

$$\begin{aligned}
CVA &= E_t[\int_t^T (r_c - \mu_c)(1 - \eta_c) I(V > 0) V_s^* e^{-\int_t^s r_e du} ds], \\
DVA &= E_t[\int_t^T (r_b - \mu_b)(1 - \eta_b) I(V \leq 0)(-V_s^*) e^{-\int_t^s r_e du} ds], \\
CFA &= E_t[\int_t^T (\mu_c - r)(1 - \eta_c) I(V > 0) V_s^* e^{-\int_t^s r_e du} ds], \\
DFA &= E_t[\int_t^T (\mu_b - r)(1 - \eta_b) I(V \leq 0)(-V_s^*) e^{-\int_t^s r_e du} ds], \\
LVA &= E_t[\int_t^T \eta_c ((1-\chi)(\mu_c - r) + \chi(r_p - r)) I(V > 0) V_s^* e^{-\int_t^s r_e du} ds], \\
&\quad - E_t[\int_t^T \eta_b ((1-\chi)(\mu_b - r) + \chi(r_p - r)) I(V \leq 0)(-V_s^*) e^{-\int_t^s r_e du} ds]
\end{aligned} \tag{16}$$

**4.1 LVA versus ColVA**

The term $(1-\chi)(\mu_c - r) + \chi(r_p - r)$ in LVA can be seen as C's effective collateral rate. $\chi(r_p - r)$ reflects primarily repo cost and leads to colVA (collateral value adjustment). For a pure receivable to B, V≥0, $r_b$ term drops out from equation (16), colVA and LVA are

$$colVA = E_t[\int_t^T \eta_c \chi(r_p - r)V_s^* e^{-\int_t^s (r_c(1-\eta_c)+\eta_c((1-\chi)\mu_c+\chi r_p))du} ds]$$
$$LVA = colVA + E_t[\int_t^T \eta_c(1-\chi)(\mu_c - r)V_s^* e^{-\int_t^s (r_c(1-\eta_c)+\eta_c((1-\chi)\mu_c+\chi r_p))du} ds] \quad (17)$$

$(1-\chi)(\mu_c - r)$ reflects funding cost of the secured but unfunded exposure due to CSA haircut being less than the repo haircut. With comingled collateral and $h_p = h_c$, $LVA=colVA$.

Our colVA formulae differs from other definitions such as Burgard and Kjaer (2013),

$$colVA = \int_t^T s_x(u)e^{-\int_t^u (r+\lambda_B(s)+\lambda_c(s))ds} E[X(u)]du \quad (18)$$

where $s_x$ is the spread earned on collateral, equivalent to our $r_L$-$r$ for cash collateral or $r_p$-$r$ for non-cash collateral, $X$ is the amount of collateral, $\lambda_B$ and $\lambda_C$ the credit spreads (thus containing the liquidity basis) of a zero recovery zero coupon bonds issued by party B and C respectively. It is easy to verify that assuming full cash collateral (same amount as the risk free fair value $V^*$), equation (18) still involves $\lambda_B$ and $\lambda_C$ which are not seen in Piterbarg (2010) and our results. Our definition of $LVA$[7] is therefore different and a bit broader than what has been shown in the literature.

With deterministic η, these XVA can be computed without recursion. We see easily that CRA (CVA+FVA) proportionally goes down as LVA goes up, due to the collateralization factor $η_c$. In the cash collateral case, $r_p$ should be replaced by $r_L$.

LVA as cost can be charged back to customers. The secured party B conducts the repo transaction and incurs the cost on C's behalf. Intuitively, the pledger (party C) could do the repo and hand over the cash to party B, without changing the trade economics. By taking in cash, party B would price the trade at OIS discount to arrive at a higher price $V^*$, implying zero LVA as $\chi=1$, and $r_p=r$ in (17). Party C, however, does not gain from being paid this higher price as his net economics has to include his repo cost which should be the same as B's LVA. All-in-all, it doesn't matter whether he hands over the security for B to repo or does the repo himself then hands over the cash, so long as both parties access the same repo market.

---

[7] Apart from these differences, we prefer the term LVA to colVA because it better reflects that collateralizing derivatives is providing liquidity on the derivatives, i.e., private financing of derivatives.

Now if the CSA allows both cash and security collateral, party C then holds a collateral option. If B prices the trade as if cash collateralized, C could realize an arbitrage benefit in the amount of LVA by posting securities instead of cash. Cash obviously has zero LVA, while government securities have non-zero, positive LVA. And non-government securities are expected to have greater LVA than government securities. To avoid collateral liquidity arbitrage, B would have to price the trade by maximizing LVA among all possible collateral posting combinations. LVA can be used to measure these collateral choices, as will be explored in the next section.

Similarly, for a pure payable to party B, $V \leq 0$, party C's funding curve drops out and LVA will be a benefit. For swap-like hybrids, LVA cost and benefit will coexist, as the first and second expressions shown in the last row of equation (16). Liquidity cost or benefit may warrant clarification. Suppose B has two identical derivatives liabilities (say short call options) dealt with two counterparties $C_1$ and $C_2$. The CSA with $C_1$ allows only full cash collateral, while $C_2$ accepts only securities. For the trade with $C_1$, $LVA_1=0$ and $V_1=V^*<0$. For the trade with $C_2$, $LVA_2<0$, and $0>V_2=V^*-LVA_2>V^*$. This means that B's liability with $C_2$ is less due to $LVA_2$, so it is a benefit for B. From $C_2$'s point of view, it incurs a cost to turn the securities to cash. Counterparty's cost is B's own benefit.

The non-recursive solution (eqt. 17) afforded by a pure asset or liability does not exist in general and one has to seek numerical solutions. The Monte Carlo simulation with regression approach designed for the switching discount rate in Lou (2016a) can be easily extended to cope with the added complexity of fractional and/or non-cash collateral.

Finally, we can derive PDE and XVAs in a similar fashion with the traditional riskless close-out. Results are skipped as this is deemed nearly obsolete, as all major financial firms have either switched to ISDA 2002 which introduces the risky close-out or adopted ISDA 2009 Close-out Protocol if they are still using ISDA 1992 which is the source of the traditional riskless close-out. Obviously the market value close-out is closer to the ISDA 2002 and ISDA Close-out 2009.

**4.2 Collateral rate inputs**

Equation (13) depends on the repo rate input. The challenge lies in the observability of repo rates and how to supply them when not observable. Even in the most liquid Treasuries repo market, quoted repo rates seldom go beyond 3 month tenor. Repo tenors extending to 1 year and beyond are not impossible nowadays, especially when repos are utilized as a short term investment vehicle (Lou 2016c). But the longest maturity of a derivatives netting set can easily exceed 10 years. In fact, netting sets are perpetual as new trades get added, until one of the parties desires to terminate the relationship or defaults. An analytical model to predict repo rates becomes the only viable solution at present time.

Treating repo as a debt product and starting from the standard Black-Scholes-Merton set-up, Lou (2016b) considers repo gap risk during a margin period of risk (MPR) and applies an innovative economic capital approach as the gap risk is neither hedgeable nor diversifiable. Gap risk pricing introduces two new adjustments, the gap risk economic value adjustment (GAP_EVA) for expected gap loss, and capital valuation adjustment (KVA) for economic capital charge. A repo break-even rate formulae is derived,

$$r_p - r = RoE \cdot E_c + \mu_0 + \lambda \cdot El \tag{19}$$

where $E_c$ is repo economic capital, $RoE$ return on equity, $\mu_0$ cost of fund, $El$ expected gap loss, and $\lambda$ borrower's hazard rate.

GAP_EVA relates to $\lambda El$, and is very small even under marginal haircuts, compared with KVA which relates to $RoE \cdot E_c$. At zero haircut, for instance, a one-year repo with 10 day MPR on US main equities could command about 50 bp KVA for a 'BBB' rated borrower. Increased haircut reduces KVA, e.g., to 4 bp at 10% haircut, while GAP_EVA is only a fraction of a basis point. $E_c$ depends on repo haircut and borrower's credit quality $\lambda$. The former as a model input can be drawn from the tri-party repo market where haircuts are relatively stable even during the financial crisis. The latter is observable in the secondary debt market and/or CDS market.

$\mu_0$ is near credit risk free assets' funding rate, e.g., the rate on the remaining balance of a loan when its 99.9-percentile value-at-risk is taken out, thus is a measure of tail financing or pure funding liquidity. Unlike similar intended issuer liquidity rate, $\mu_0$ is a broad market measure, not specific to a security but possibly asset class specific. For our purposes, we take the Libor-OIS swap spread curve as a proxy and leave its proper identification to future research.

When forecasting repo rates beyond 1 year, GAP_EVA should remain small and can be ignored as the duration of credit risk remains within the MPR in days. KVA charge is stable, if $\lambda$ is flat, as the computation of $E_c$ employs stressed historical data. Otherwise, forward $\lambda$ could be used. $\mu_0$ can incorporate Libor-OIS spread curve term structure. Once the effective discount rate is determined for the entire duration of the netting set, we can proceed to compute LVA and explore collateral optimization.

## 5. Application in Collateral Optimization

As derivatives collateralization has boomed, reliable and efficient collateral management becomes a critical business function. The industry has coined the term *collateral optimization*[8] to mean the procedure of finding the right collateral that meets each CSA's collateral eligibility criteria. Roughly, if a CSA allows BBB rated corporate bonds, the procedure would exhaust 'BBB' rated bond holding before sending out 'A' rated bonds. Some intuitive haircut arbitrage is included. For example, if party B's CSA sets 5% haircut on US Treasuries while C's has 0%, it would prefer sending the Treasuries to C to B.

Such strategies recognize that there is value in the form of cheapest delivery collateral, apart from more prominent embedded CSA optionality such as currency delivery choices and other significant structural asymmetries and imperfection (Fujii and Takahashi, 2013). Lacking a quantitative measure, they may not capture collateral cost and benefit, which is what LVA is designed for. Below we develop a linearized procedure to optimize derivatives collateral management through LVA.

### 5.1 Single netting set
When collateral is segregated, repo financing cost does not enter LVA as $\chi=0$ in equation (17). A party called upon for collateral can simply post securities with repo haircuts that exceed CSA haircuts by the most. The scope of our collateral optimization is therefore limited to comingled securities, i.e., $\chi=1$.

---

[8] Per ISDA (2015) margin survey, 83% of large firms optimize collateral, and most operate systemically on a daily basis.

A particular asset's LVA per unit can be computed by first assuming that it has unlimited quantity such that it could completely fulfill a firm's CSA posting requirement, then normalizing so obtained LVA with the required quantity. Let $e$ denote the LVA per unit of security priced at B, $e = B(1-h_c)LVA/V$.

Collateral is necessarily posted at the netting set level. For non-cash collateral, oftentimes the collateral asset portfolio consists of many different assets. $h_c$, $h_p$ and $r_p$ in equations (14 & 17) are indeed effective CSA haircut, repo haircut, and repo rate. Specifically, if $m$ assets are posted under a netting set, each with market value of $A_i$ and repo market haircut $h_{pi}$ and repo rate $r_{pi}$, $i=1, 2, \ldots, m$, we have the following,

$$\bar{\chi} = \sum_i w_i \frac{(h_{pi} - h_{ci})^+}{1 - h_{ci}}$$
$$S_{pi} = (1 - \frac{(h_{pi} - h_{ci})^+}{1 - h_{ci}})(r_{pi} - r) \qquad (20)$$
$$(1 - \bar{\chi})(r_p - r) = \sum_i w_i S_{pi}$$

where $w_i$ is the fraction of the exposure covered by $i$-th asset, $w_i = \frac{(1-h_{ci})A_i}{L}$, $S_{pi}$ adjusted repo rate, and $\chi = 1 - \bar{\chi}$.

For a single counterparty, when different securities or classes of securities are available, a pecking order can be established in term of the LVA per unit and securities would be sent out of the door in that order.

### 5.2 Multiple netting sets

The scheme gets more interesting when multiple counterparties are involved, as LVA could depend on counterparty's credit and its netting set characteristics. In the simplest case, suppose that party C has identical CSA eligible collateral and associated haircuts with a number of dealers and has a netting set of pure payables with each party. LVA is then only liability-side (here party C) dependent, per equation (17). C is indifferent as to which dealer party to post first. Subsequently, the procedure becomes a haircut arbitrage exercise. In fact, from equation (20), what matters is the adjusted repo spread $S_{pi}$. When repo haircuts are less than or equal to collateral haircuts, securities with higher repo rates should be sent out first; otherwise, $S_{pi}$ becomes $\frac{1-h_{pi}}{1-h_{ci}}(r_{p_i}-r)$ which should be used as the basis to remit securities. This in fact reflects the essence of typical optimization procedures.

For netting sets consisting of swap like hybrid assets (or liabilities), the discount rate involves both counterparty's rates. A security may not have the same utility as each netting set's maturity profile and moneyness are different. Especially considering the other party (B) also has a CSA posting option when it becomes the liability-side in some future time. C has to make assumptions about what party B might post. Because a bilateral CSA is usually symmetric in terms of eligible securities and their haircuts, without prior or private knowledge of how B's collateral pool looks like, the only reasonable assumption to make is that B's collateral pool mirrors C's. In fact, equation (16) has implied such an assumption by using the same $\chi$ on both LVA cost and benefit.

Suppose the firm has a collateral pool of $M$ assets and $N$ counterparties with bilateral full CSA. Let $B_i$ and $Q_i$ denote the market price and holding quantity of $i$-th asset, and $V_j$ the mark-to-market exposure of $j$-th counterparty. Assuming $V_j$ to be met with unlimited quantity of $i$-th asset, we proceed to find LVA, label it as $LVA_{ij}$ and normalize it to obtain $e_{ij}$. A collateral posting scheme is an allocation $\{q_{ij}\}$, where $q_{ij}$ is the quantity of the $i$-th collateral asset allocated to the $j$-th netting set, $Q_i \geq q_{ij} \geq 0$, such that we maximize the total LVA,

$$\max(\sum_{i,j} q_{ij} e_{ij})$$

subject to the following linear inequality constraint and equality constraint:

$$\sum_j q_{ij} \leq Q_i,$$
$$\sum_i q_{ij}(1-h_{p_i})B_i = V_j. \tag{21}$$

Once the unit LVAs are computed, standard linear programming techniques can be employed. A convenience is to always allow one of the assets to be cash with zero haircut, price of 1 and arbitrarily large quantity, even if one has sufficient assets to cover all demands. The pool to be allocated should consist all non-segregated collateral posted by counterparties. Eligibility criteria, such as rating restriction, concentration and price limit etc., have to be run in a separate process. Securities found to be ineligible for j-th netting set can be incorporated into the LP scheme by setting the lower and upper bounds of $q_{ij}$ to zeros.

A collateral security, if qualified as a high quality liquid asset (HQLA), can also be counted towards BASEL III liquidity coverage ratio (LCR). Since this is a regulatory requirement, we can build it as an additional constraint. Rewrite the inequality in equation (21) in a slack form with nonnegative slack variables $s_i$, $s_i \geq 0$,

$$\sum_j q_{ij} + s_i = Q_i,$$
$$\sum_i s_i (1 - h_{L_i}) B_i \geq H \tag{22}$$

where $H$ is the required (HQLA) coverage value (product of LCR and the contingent cash flow during a 30 day period) and $h_{Li}$ is the applicable LCR haircut for $i$-th asset.

Note for netting sets of positive MTM, party C does not post. Collateral pledged by other parties, if any and rehypothecatable, can be simply added to the available pool of eligible collateral. The same allocation process follows.

In the above, we have handled full collateralization with non-cash collateral. Partial collateralization does exist, although to a much less extent. A non-zero threshold, for instance, creates a pocket of fixed uncollateralized exposure, while leaving the rest collateralized. The procedure can be a good approximation as the portion of uncollateralized exposure does not change much (only to the extent the discount rate is effected) while the collateralized portion is being optimized.

The proposed optimization procedure is a simplification of the complex collateral management process that involves both qualitative and quantitative aspects, some of which are non-linear in nature. We have in fact linearized the LVA objective function. As the system's dimension is large, considering the number of counterparties and the number of collateral securities, some sort of linearization and simplified modeling are necessary in order to obtain a practical solution. The advantage of an LP based procedure is its ease of incorporating firm specific features through additional constraints, e.g., initial margin requirement. The LP problem formulated and the repo rate formulae can still apply when other definitions of LVA or colVA are used instead of our definition, so that it hopefully could form a basis for full-fledged collateral optimization procedures.

**6. Discussions and Results**

To show impact of collateralization, we calculate CRA, LVA and XVA of a European at-the-money call option. As shown in Figure 1, at 100% collateralization, CRA is zero for both long and short positions. At zero collateralization LVA (as the difference between XVA and CRA, not directly shown) is zero so that XVA=CRA. For the long call, CRA as the sum of CVA and CFA is the largest at zero collateralization (i.e., uncollateralized). As collateralization increases, CRA reduces while LVA increases. At full collateralization, CRA is zero and LVA is at its largest. The gap between the long XVA and the short XVA becomes the bid and ask spread of this standalone call option.

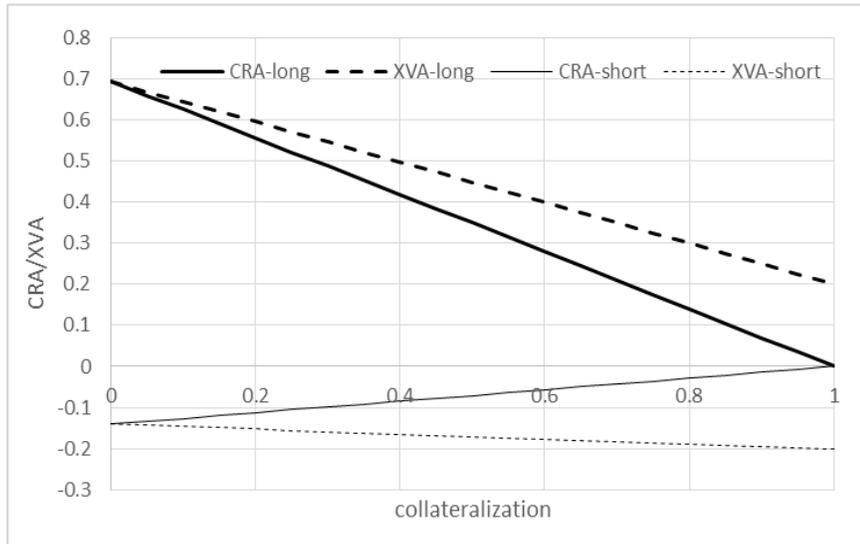

Figure 1. Comparisons of CRA and XVA for a long call option and a short call option, S=K=100, vol=50%, T=1 year, r=1%. Party B credit spread to OIS 1.25%, party C 3%, collateral rate 1%.

Next, we compute XVA for three sample portfolios of 1000 swaps, randomly generated with uniform maturity distribution from 0.25 to 30 years, and swap rates around at-the-money rate with 10%, 50% and 90% payer population. Party B's and C's credit spreads used are both 125 bp, close to 100 bp collateral repo spread, so total XVA decreases rather slowly from uncollateralized (collateralization=0) to full collateralization, see Figure 2. LVA increases from zero when collateralization 0 and reaches maximum at collateralization 1, while CRA is the reverse, from maximum to zero.

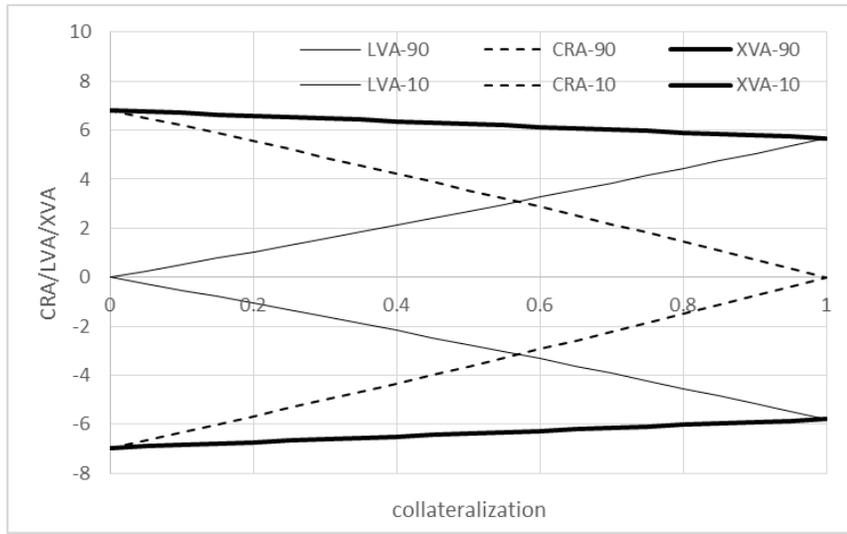

Figure 2. CRA/LVA/XVA for sample one-sided swap portfolios. Top half shows a random portfolio of 10% payer and the bottom half shows 90% payer.

Table 1. XVA decomposition into LVA and CRA (=CVA-DVA+CFA-DFA), for the three test swap netting sets when uncollateralized, partially (50%) collateralized and fully collateralized.

|  | No collateral | | | Half collateral | | | Full collateral | | |
| --- | --- | --- | --- | --- | --- | --- | --- | --- | --- |
|  | 90% | 50% | 10% | 90% | 50% | 10% | 90% | 50% | 10% |
| NPV | -71.55 | 1.77 | 70.04 | -72.13 | 1.80 | 70.61 | -72.72 | 1.83 | 71.19 |
| XVA | -6.96 | 0.37 | 6.82 | -6.38 | 0.34 | 6.25 | -5.79 | 0.31 | 5.67 |
| LVA | 0.00 | 0.00 | 0.00 | -2.75 | 0.14 | 2.69 | -5.79 | 0.31 | 5.67 |
| CRA | -6.96 | 0.37 | 6.82 | -3.63 | 0.19 | 3.56 | 0.00 | 0.00 | 0.00 |
| CVA | 0.07 | 0.25 | 4.55 | 0.04 | 0.13 | 2.34 | 0.00 | 0.00 | 0.00 |
| DVA | 4.65 | 0.00 | 0.07 | 2.39 | 0.00 | 0.04 | 0.00 | 0.00 | 0.00 |
| CFA | 0.04 | 0.12 | 2.37 | 0.02 | 0.07 | 1.27 | 0.00 | 0.00 | 0.00 |
| DFA | 2.42 | 0.00 | 0.04 | 1.30 | 0.00 | 0.02 | 0.00 | 0.00 | 0.00 |

As mentioned earlier, CRA can be further decomposed into bilateral CVA and FVA. Table 1 below shows the decomposition of swap portfolio XVA into CRA and LVA with the former further split into CVA/DVA and FCA/DFA for zero, 50% and 100% collateralization. All XVAs are expressed in a running spread in basis points. As expected, the 90% payer portfolio is dominated by DVA and DFA while the 10% payer portfolio is by CVA and CFA. As far as LVA is concerned, for the 90% payer swap portfolio (payable), LVA shows a benefit of 5.79 bp, and for the 10% payer swap portfolio as a receivable, there is a cost of 5.67 bp.

CSAs commonly have an option to post cash or Treasuries. Market participants typically treat them with the OIS discounting regardless of the haircuts applied to the Treasuries. Strictly speaking, there is an LVA involved due to the difference between Treasuries repo rates and the Fed funds rate. Post the financial crisis, the 3 month Treasuries GCF repo rate has been on average about 10 bp higher than 3 month OIS rate. Table 2 shows LVA for the three sample swap portfolios with $r_L-r=0.1\%$. The two imbalanced portfolios are of about 1.25 bp of LVA. Obviously these will multiply when GCF and OIS spread widens. For instance, late October and early November 2016 saw the spread at 25 bp. Dealer banks therefore can't simply ignore the difference for good and take for granted to apply OIS discounting to full CSA that allows non-cash collateral.

Table 2. LVA due to OIS and treasuries repo rate differences (10 bp spread) for sample interest rate swap portfolios.

| Portf | 90% | 50% | 10% |
|---|---|---|---|
| npv | -77.26 | 2.07 | 75.63 |
| LVA | -1.25 | 0.07 | 1.22 |

To demonstrate the use of LVA for derivatives collateral management, we give an example of party C having four full CSA netting sets with four different credit quality counterparties, hypothetically rated 'AA', 'A', 'BBB', 'BB', with 5 year CDS at 125, 250, 500 and 1000 bp respectively. His collateral pool consists of 6 asset classes, including 10 and 30 year US Treasuries, S&P 500 main equities, 'A' rated corporate bonds with 5 to 10 years of remaining maturity, and 'AAA' and 'AA' rated commercial mortgage backed securities (CMBS) with 5 to 10 years remaining maturity. Each asset class's available market value is set to 75 fixed.

Economic capital is calculated with consideration of banks' credit quality per asset class, using historically estimated double exponential jump diffusion process (Lou 2016c), see Table 3. Sample CSA and repo haircuts are also shown. The cost of fund is taken from LIBOR-OIS spread, roughly at 10 bp at the short end (3 month) and about 50 bp at 30 years.

Table 3. Repo economic capital (%) of selected asset classes for four credits hypothetically rated 'AA', 'A', 'BBB', and 'BB', under given haircuts.

| \Cpty Rtg | AA | A | BBB | BB | CSA HC | Repo HC |
|---|---|---|---|---|---|---|
| UST_10y | 0.08 | 0.17 | 0.4 | 0.8 | 0.02 | 0.03 |
| UST_30y | 1.2 | 1.7 | 2.19 | 2.7 | 0.04 | 0.03 |
| S&P_500 | 1.61 | 2.53 | 3.41 | 4.28 | 0.15 | 0.075 |
| CMBS_AAA5y | 0.32 | 0.69 | 1.49 | 2.41 | 0.12 | 0.06 |
| CMBS_AA5y10 | 1.15 | 2.4 | 3.89 | 5.5 | 0.18 | 0.075 |
| Corp_A5y10 | 0 | 0.01 | 0.02 | 0.04 | 0.09 | 0.05 |

The four netting sets are randomly generated interest rate swap portfolios, each with 1000 swaps of 90%, 80%, 70%, and 60% payers, and having negative (risk-free) mark-to-markets of -118.007, -90.641, -60.98, and -29.915 respectively. To allocate collateral, one would need MTM liabilities first but that depend on collateral posted. A collateral posting scheme therefore is necessarily iterative. Our scheme starts with the OIS discounted MTM, then computes each asset class's unit LVA per counterparty, e.g., as listed in Table 4, assuming C is an 'A' rated firm.

Table 4. Unit LVA of each asset per counterparty assuming party C is 'A' rated.

| \Cpty Rtg | AA-set | A-set | BBB-set | BB-set |
|---|---|---|---|---|
| UST_10y | 0.0441 | 0.0439 | 0.0424 | 0.0414 |
| UST_30y | 0.0587 | 0.0584 | 0.0564 | 0.0551 |
| S&P_500 | 0.0665 | 0.0660 | 0.0638 | 0.0623 |
| CMBS_AAA5y | 0.0492 | 0.0489 | 0.0472 | 0.0460 |
| CMBS_AA5y10 | 0.0653 | 0.0648 | 0.0625 | 0.0609 |
| Corp_A5y10 | 0.0425 | 0.0423 | 0.0409 | 0.0400 |

Linear programming solver is quite standard. We use Matlab's *linprog* function for large scale programs. The initial allocation results are shown in Table 5.

Table 5. Party C's initial collateral asset allocation per counterparty's netting set.

| \Cpty Rtg | AA-set | A-set | BBB-set | BB-set |
|---|---|---|---|---|
| UST_10y | 0 | 0 | 0 | 0 |
| UST_30y | 1.15 | 68.85 | 0 | 0 |
| S&P_500 | 70 | 0 | 0 | 0 |
| CMBS_AAA5y | 0 | 27.9 | 42.1 | 0 |
| CMBS_AA5y10 | 70 | 0 | 0 | 0 |
| Corp_A5y10 | 0 | 0 | 26.3 | 32.87 |

Now that we have an allocation, each netting set's LVA can be recalculated and the allocation updated. Table 6 shows the first update of the allocation. Last row shows updated MTM for each netting set. Subsequent updates yield little improvement and results are omitted here.

Table 6. Updated collateral asset allocation.

| \Cpty Rtg | AA-set | A-set | BBB-set | BB-set |
|---|---|---|---|---|
| UST_10y | 0 | 0 | 0 | 0 |
| UST_30y | 0 | 70 | 0 | 0 |
| S&P_500 | 61.37 | 8.63 | 0 | 0 |
| CMBS_AAA5y | 0 | 12.45 | 57.55 | 0 |
| CMBS_AA5y10 | 70 | 0 | 0 | 0 |
| Corp_A5y10 | 0 | 0 | 8.18 | 31.47 |
| Updated MTM | -109.57 | -85.49 | -58.10 | -28.64 |

**7. Conclusions**

Imperfect collateral exists when collateral is segregated or non-cash. We show that the effective derivative financing rate is a weighted sum of the bond curve, the liquidity rate, and the repo rate with the weights determined as proportions of unsecured exposure, secured yet unfunded exposure, and repo funded exposure. The effect of private financing of derivatives via collateralization is generally non-linear as the discount rate switches between two counterparties to stay on the liability-side.

Non-overlapping valuation adjustments are introduced based on a natural decomposition of a precisely defined total valuation adjustment. The portion associated with the bond curve is counterparty risk adjustment (CRA) which can be further decomposed into CVA and FVA, corresponding to the default risk premium and the funding basis of the bond credit spread. The portion associated with collateral is defined as liquidity value adjustment (LVA) which consists of a part associated with the liquidity rate attributed to haircut differences, and another part associated with the securities' repo financing cost defined as collateral value adjustment (colVA).

LVA thus depends on both repo haircuts and repo rates, providing a theoretical construct linking pricing of privately financed derivatives to the securities financing market. To address the tenor mismatch between perpetual CSA and repo tenors of a few months, the break-even repo formulae is employed to estimate haircut dependent long term repo rates. We find that LVA for government security collateralized swap portfolios could reach a few basis points. Applying OIS discounting to fully collateralized netting sets irrespective of collateral nature therefore warrants discretion.

By treating LVA as a quantitative measure of collateral value and computing it for each counterparty and every collateral asset, a linear programming problem can be formulated to maximize total LVA that could serve as the core of a full-fledged collateral optimization procedure. Basel's leverage coverage ratio and other enterprise features can be incorporated as constraints. Numerical examples show that LVA could be sizable for long average duration, deep in or out of the money swap portfolios, especially when the repo market is in an expansion cycle with low haircuts.

**References**

Anderson, L., D. Duffie, and Song (2016), Funding Value Adjustment, ssrn preprint.

Brigo, D., Q. Liu, A. Pallavicini, and D. Sloth, Nonlinear Valuation under Collateral, Credit Risk and Funding Costs: A Numerical Case Study Extending Black-Scholes, Handbook in Fixed-Income Securities, Wiley, 2014.

Burgard, C. and M. Kjaer (2011), Partial Differential Equation Representations of Derivatives with Bilateral Counterparty Risk and Funding Costs, J of Credit Risk, Vol 7, No. 3.

Burgard, C. and M. Kjaer (2013), Funding Strategies, Funding Costs, Risk, December 2013, pp 82-87.

Duffie, D. and M. Huang (1996), Swap Rates and Credit Quality, J of Finance, 51(3), pp 921-949.

Fujii, M. and A. Takahashi (2013), Derivative Pricing under Asymmetric and Imperfect Collateralization and CVA, Journal of Quantitative Finance, 13(5), pp 749-768.

Hull, J. and A. White (2014), Collateral and Credit Issues in Derivatives Pricing, J of Credit Risk, 10(3).

ISDA (2015), ISDA Margin Survey 2015, August 2015.

Johannes, M. and S. Sundaresan (2007), The Impact of Collateralization on Swap Rates, J. of Finance, Vol 62(1), pp 383-410.

Lou, Wujiang (2015), CVA and FVA with Liability-side Pricing of Derivatives, *Risk*, August 2015, pp 48-53.

Lou, Wujiang (2016a), Liability-side Pricing of Swaps, *Risk*, April, pp 66-71.

Lou, Wujiang (2016b), Gap Risk KVA and Repo Pricing, *Risk*, November, pp 70-75.

Lou, Wujiang (2016c), Repo Haircut and Economic Capital, SSRN preprint, Feb 2016.

Piterbarg, V. (2010), Funding Beyond Discounting: Collateral Agreements and Derivatives Pricing, *Risk* February, pp 97-102.

Piterbarg, V. (2012), Cooking with Collateral, *Risk* August, pp 46-51.